\documentclass[prd,nofootinbib,twocolumn]{revtex4}

\usepackage{graphics}
\usepackage{bm}
\usepackage{latexsym,amssymb,amsmath}

\begin{document}

\title{
A Classical Treatment of Island Cosmology
}

\author{
Sourish Dutta
}
\affiliation{
CERCA, Department of Physics, Case Western Reserve University,
10900 Euclid Avenue, Cleveland, OH 44106-7079, USA.}

\begin{abstract}
\noindent
Computing the perturbation spectrum in the recently proposed Island Cosmology remains an open problem. In this paper we present a classical computation of the perturbations generated in this scenario by assuming that the NEC-violating field behaves as a classical phantom field. Using an exactly-solvable potential, we show that the model generates a scale-invariant spectrum of scalar perturbations, as well as a scale-invariant spectrum of gravitational waves. The scalar perturbations can have sufficient amplitude to seed cosmological structure, while the gravitational waves have a vastly diminished amplitude. 
\end{abstract}

\maketitle
\section{Introduction}
\label{intro}

Recent cosmological obervations \cite{CMBresults1} seem to be consistent with the hypothesis that we live in a flat $\Lambda$-dominated Universe seeded by primordial fluctuations that were predominantly adiabatic and nearly scale-invariant. The most successful class of models leading to such a Universe are the Inflationary  models \cite{guthinflation} (see, for instance \cite{riotto}, \cite{brandenberger1}, \cite{liddlelyth} and \cite{langlois} for recent reviews). 

Recently a new cosmological model has been proposed in Ref. \cite{islands} called Island Cosmology. In this scenario, the presently observed state of the Universe (i.e., inflating with a cosmological constant), is considered to be the eternal state. Large-scale explosive events which involve local violations of the Null Energy Condition (NEC) lead to the formation of ``islands'' of matter, one of which is our observed universe. The cause of these explosive events is attributed to quantum fluctuations in some field. While a complete calculation of density perturbations in this scenario (which would require a proper treatment of the back reaction in quantum gravity) is not attempted, the authors show that any other free scalar field responding to the metric fluctuation would produce a scale-invariant spectrum.
 
In this paper we consider a special version of Island Cosmology where we model the behavior of matter during the NEC-violating event as a classical phantom field in order to compute the perturbation spectrum. The different stages of our model are described below (see  Fig. \ref{fig1}):
\begin{figure}
\scalebox{0.20}{\includegraphics{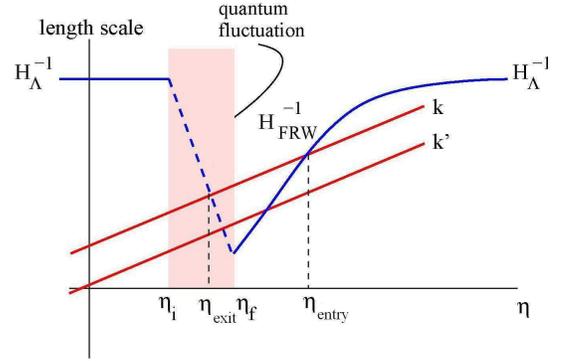}}
\caption{\label{fig1} Sketch of the behavior of the Hubble length 
scale with conformal time, $\eta$, in the Island model, and 
the evolution of fluctuation modes. At early times, inflation is 
driven by the presently observed dark energy, assumed to be a cosmological 
constant. As the cosmological constant is very small, the Hubble length
scale is very large -- of order the present horizon size. 
Exponential inflation in some horizon volume ends not due to the
decay of the vacuum energy as in inflationary scenarios but due
to a quantum fluctuation in the time interval $(\eta_{i}, \eta_{f})$ 
that violates the null energy condition (NEC). The NEC violating
quantum fluctuation  causes the Hubble length scale to decrease. 
After the fluctuation is over, the universe enters radiation dominated 
FRW expansion, and the Hubble length scale grows with time. The 
physical wavelength of a quantum fluctuation mode starts out less 
than $H^{-1}_\Lambda$ at some early time $\eta_i$. The mode exits 
the cosmological horizon during the NEC violating fluctuation 
($\eta_{\text{exit}}$) and then re-enters the horizon at some later epoch 
($\eta_{\text{entry}}$) during the FRW epoch (The modes are drawn as straight lines for illustrative purposes only, they actually grow in proportion to the scale factor).
}
\end{figure}
\begin{enumerate}
\item {\emph{The de Sitter Phase:}} We start with the assumption that the initial state of the Universe is de Sitter space inflating with the currently observed value of the Hubble constant. 
\item {\emph{The Phantom Phase:}} A quantum fluctuation in some field (chosen to be a light scalar field for simplicity) in the expanding phase of de-Sitter spacetime,   and occuring over a Horizon-sized volume drives the Hubble constant to a high value. Such a fluctuation necessarily violates the Null Energy Condition within this volume. In this treatment we make the working hypothesis that during this phase the energy content within a (current) Hubble sized volume of the Universe behaves like phantom energy, i.e., a classical perfect fluid with equation of state 
$w=p/\rho<-1$, where $p$ and  $\rho$ stand for pressure and density respectively, and $\rho>0$. The Hubble length decreases sharply during this period, though the Universe keeps expanding. 
\item {\emph{The FRW Phase:}} After the explosive event, the Hubble constant is large, and classical radiation fills the volume. Rapid interactions thermalize this radiation, and this part of the Universe then follows a radiation-dominated FRW evolution
leading to the Universe that is observed today. 

\item {\emph{The Dilution Phase:}} Eventually cosmic expansion dilutes out this aggregation of matter, and the volume of space is restored to its initial de Sitter state.  
\end{enumerate}
In the remainder of this paper we will show that this model predicts a (nearly) scale invariant spectrum of scalar perturbations with amplitude sufficient to seed structure, as well as a scale invariant spectrum of tensor perturbations of considerably smaller amplitude. 

A similar situation was studied by Y.S. Piao in Ref. \cite{piao} using a different calculational approach.

The layout of the paper is as follows: Section \ref{The Model} describes the model in greater detail. Section \ref{Calculational Strategy} discusses the assumptions and outlines our calculational strategy. Section \ref{The Perturbation Spectrum} presents a detailed calculation of both the scalar and tensor power spectra, and Section \ref{Conclusions} summarizes our findings.

\section{The Model}
\label{The Model}
In the following subsections, we describe in more detail the three stages of the
model, stating the assumptions involved in each stage. 

We set our notation as follows: let us choose our cosmic time coordinate $t$ such that the 
phantom phase (described in Section \ref{The Phantom Phase}) begins at $t=t_{i}=0$. In this paper we find it more convenient to work in conformal time ($\eta$) where $dt=a(\eta) \,d\eta$ and $\eta$ ranges between $(-\infty\text{,}\ 0)$ as $t$ goes from $-\infty$ to $+\infty$. We set our conformal time coordinates such that the period of phantom cosmology lasts between $\eta=\eta_{i}=-1/H_{\Lambda}$ 
and $\eta=\eta_{f}$. The above choices of $t_{i}$ and $\eta_{i}$ are arbitrary and for convenience. Primes denote derivatives with respect to 
conformal time, and dots denote derivatives with respect to cosmic time. We adopt the convention that the suffix $i$ denotes the value of a quantity at $\eta=\eta_{i}$ 
and the suffix $f$ denotes the value at $\eta=\eta_{f}$.
\subsection{The de Sitter phase}
\label{thedesitterphase}
This phase represents the initial state of the Universe, before the onset of the 
phantom behavior. In this phase, we assume that the Universe is de Sitter space inflating due to the observed dark energy, which we assume is a cosmological
constant. The Hubble parameter ($H$) has the same value that it has today, which we call $H_{\Lambda}$ . This expanding de Sitter background can be part of a classical de Sitter spacetime with no
beginning and no end, with early contraction and then
expansion. We will only consider the expanding phase of
the de Sitter spacetime in the following discussion. 

We assume that the matter content of the Universe is a classical scalar field ($\phi$) having a Lagrangian $\cal L$ given by:

\begin{equation}
\label{Lagrangian}
{\cal L}=\lambda\partial^{\mu}(\phi) \partial_{\mu}(\phi)-V(\phi)
\end{equation}
and a stress tensor given by:
\begin{equation}
\label{Stress Tensor}
T_{\mu\nu}=\lambda\partial_{\mu}(\phi)\partial_{\nu}(\phi)-g_{\mu\nu}{\cal L}
\end{equation}
$g^{\mu\nu}$ represent the components of the metric, and $V(\phi)$ represents the potential.
For the sake of generality, we have inserted the constant $\lambda$ which 
determines the sign of the kinetic term. Obviously, $\lambda=+1$ for an ordinary 
classical scalar field, and $\lambda=-1$ for a phantom field.

The equation of motion of the field is the Klien Gordon equation:
\begin{equation}
\label{KG}
\lambda\frac{1}{\sqrt{-g}}\partial_{\mu}\left( \sqrt{-g} g^{\mu\nu}\partial_{\nu}\phi\right)
+\frac{\partial V}{\partial \phi}=0
\end{equation}


The scale factor $a(t)$ and Hubble value $H(t)$ during this period  can be written as follows:

For $(-\infty<t\leq t_{i}= 0)$:

\begin{eqnarray}
a(t)&=&e^{H_{\Lambda} t}\\
H(t)&=&\frac{\dot{a}(t)}{a(t)}=H_{\Lambda}
\end{eqnarray}

 In terms of conformal time, for $(-\infty<\eta\leq \eta_{i})$:

\begin{eqnarray}
a(\eta) &=&-\frac{1}{\eta H_{\Lambda}}\\
{\cal H}(\eta )&=&\frac{a'(\eta)}{a(\eta)}=-\frac{1}{\eta}
\end{eqnarray}


\subsection{The Phantom Phase}
\label{The Phantom Phase}In this phase, lasting between the times $\eta_{i}$ and $\eta_{f}$, the Universe undergoes an NEC-violating quantum fluctuation. We model this phase by assuming that the matter content of the Universe behaves like a phantom field $\phi_{p}$ for the duration of the fluctuation. 


We assume that this hypothetical phantom field $\phi_{p}$ is classical, i.e., its Lagrangian and stress tensor are given by Eq.~\eqref{Lagrangian} and Eq.~\eqref{Stress Tensor}, and it satisfies the Klien Gordon equation Eq.~\eqref{KG}.

Using the above equations, one can readily determine the pressure $p$ and energy density $\rho$ of the field. These work out to:
\begin{eqnarray}
p=&\frac{1}{3}\sum T_{ii}&=\lambda\frac{\phi_{p}'^{2}}{2}-a^{2}(\eta ) V(\phi_{p})\\
\rho =&T_{00}&=\lambda\frac{\phi_{p}'^{2}}{2}+a^{2}(\eta ) V(\phi_{p})
\end{eqnarray}

Clearly, $p+\rho=\lambda\phi_{p}^{2}$ indicating, as one would expect, that the NEC is violated
 during this period if our matter field is phantom $(\lambda=-1)$.

 
 Also during this phase, the Hubble horizon size drops from $H^{-1}_{\Lambda}$ to $H^{-1}_{f}$. We assume that this 
 drop is linear in cosmic time, ending at time $t_{f}$. The validity of this assumption, as well as its implications on the matter content of the Universe are discussed later. Thus,

 \begin{equation}
 \label{alphadef}
 H^{-1}(t)=H^{-1}_{\Lambda}-\alpha t\  \text{(for $0\leq t\leq t_{f}$)}
 \end{equation}
 
 Here $\alpha$ is a dimensionless parameter measuring the rate at which the horizon size changes during the 
 phantom phase. We assume that the quantum fluctuation is very abrupt, and hence $\alpha$ is very large. 

Using the definition of conformal time, it is easy to deduce that during this phase ($\eta_{i}\leq \eta \leq \eta_{f}$),
 \begin{eqnarray} 
 \label{scalefactor} a(\eta) &=& [-\alpha - H_{\Lambda}(1+\alpha) \eta ]^{-\frac{1}{1+\alpha}}\\ 
 \label{Hamba} {\cal H}(\eta) &=& \frac{a'(\eta)}{a(\eta)}=\frac{H_{\Lambda}}{[-\alpha - H_{\Lambda}(1+\alpha) \eta ]}\\ 
\label{H} H(\eta) &=& \frac{{\cal H}(\eta)}{a(\eta)}=H_{\Lambda}\left[a(\eta)\right]^{\alpha}
 \end{eqnarray}

 We also need to address the nature of the back-reaction of matter on geometry. Let us make the working hypothesis
 that the back-reaction is fully described by  the Friedmann equation.

 The actual time duration of this phase can be calculated by demanding continuity of the Hubble value at 
 $t=t_{f}$ (or equivalently, $\eta=\eta_{f}$).
 Thus we require (see Eq.~\eqref{H}),
 \begin{equation}
 \label{hequalsh}H(\eta_{f})=H_{f}
 \end{equation}
 Solving the above equation for $\eta_{f}$, we obtain
 \begin{equation}
 \eta_{f}=\frac{1}{H_{\Lambda}(1+\alpha )}\left[-\left(  \frac{H_{\Lambda}}{H_{f}} \right) ^ {1+\frac{1}{\alpha}}-\alpha\right]
 \end{equation}
 We assume that $H_{f}\gg H_{\Lambda}$ and since $\alpha\rightarrow\infty$, the first term in the square brackets can be ignored leaving us
 \begin{equation}
 \label{eta_f}
 \eta_{f}\simeq -\frac{\alpha}{H_{\Lambda}(1+\alpha)}-O\left[\frac{1}{(1+\alpha)H_{f}}\right]
 \end{equation}
 From this we can compute the duration of the phantom phase in conformal time ($\Delta\eta$), as follows: 
 \begin{align}
 \Delta\eta& =\eta_{f}-\eta_{i} &\nonumber\\
 \label{timeduration}& =\frac{1}{H_{\Lambda}}\left(\frac{1}{\alpha+1}\right)&
 \end{align}
Again, since $\alpha\rightarrow\infty$, this is a vanishingly small interval.

 \subsection{The Radiation Dominated FRW Phase}
 \label{theradiationdominatedphase}
 In this epoch we have a volume of space of Hubble length $H_{f}^{-1}$ filled with classical radiation.
 Rapid interactions thermalize the radiation, after which this volume follows a standard FRW evolution.
 
  The scale factor in this radiation-dominated epoch can therefore be written as 
\begin{equation}
 a(t)=a_{f} \sqrt{\frac{t}{t_{f}}} 
\end{equation}
In terms of the conformal time, this reduces to
\begin{equation}
\label{afrw}
a(\eta )=a_{f}+a_{f}^{2}H_{f}(\eta-\eta_{f})
\end{equation}

\section{Calculational Strategy}
\label{Calculational Strategy}

We make two key assumptions to facilitate our calculation, which we discuss below. These are:
\begin{enumerate}
\item \label{phant}During the NEC-violating explosive event the energy content of the Universe behaves as a phantom field.
\item\label{slope}During the NEC violation, the drop in the Hubble scale is linear in cosmic time.
\end{enumerate}

Assumption (\ref{phant}) is an attempt to model the behavior of the matter field during the  NEC-violating event. To calculate density fluctuations due to fluctuations in the NEC-violating field, one needs a suitable model for the evolution of the field itself during the NEC-violating fluctuation. This evolution is quantum and not described as a solution to some classical equation of motion. For the purpose of this calculation, we have made the simplifying assumption that the matter field behaves in the same manner as a classical object that would also violate the NEC and produce the same effect on the spacetime. Of course this purely classical treatment cannot substitute for a rigorous quantum mechanical treatment of the NEC-violation, but we hope that it captures the essential elements of the physics involved. 

Assumption (\ref{slope}) can be justified considering that the drop in the Hubble length need not be linear throughout the 
explosive event, but only during the window of time $\delta t$ that it takes for the scales observed today to leave the horizon.
Since the fluctuation itself is very short lived, the drop in $H^{-1}(t)$ over $\delta t$ can be well approximated to be linear. 

We now turn our attention to computing the spectrum of perturbations that would be generated in this cosmological model.
Our plan of action is the following: 
\begin{enumerate}
\item Working in $k$ (momentum or wavenumber) space, we first find expressions for the  
Mukhanov variable $v_{k}(\eta)$ 
in all the three stages of the model. The Mukhanov variable is a gauge-invariant linear combination of 
matter and metric perturbations representing the true dynamical degrees of freedom of the system. It is fully described in \cite{Mukhanov}.
\item The unknown coefficients that arise in the above expressions are then determined by demanding 
continuity of $v_{k}(\eta)$ and its time derivative $v_{k}'(\eta)$ at transition times $\eta_{i}$ and $\eta_{f}$. 
\item The adiabatic density perturbation responsible for structure in the Universe is conveniently
characterized by the curvature perturbation $\cal R$ seen by comoving observers. Once we fully determine $v_{k}(\eta)$ in the radiation dominated phase, we obtain the co-moving curvature
perturbation spectrum from the relation (see, for example, \cite{liddlelyth})
\begin{equation}
\label{powerspectrum}{\cal P}_{\cal R}(\eta )=\frac{k^{3}}{2\pi^{2}}\left\vert\frac{v_{k}(\eta )}{z(\eta)}\right\vert_{\text{FRW}} ^{2}
\end{equation}
$z$ being defined by the relation
\begin{equation}
\label{z}
z=a\frac{\phi'}{\cal H}
\end{equation}
\end{enumerate}

 \section{The Perturbation Spectrum}
 \label{The Perturbation Spectrum}

In the next three subsections (\ref{de sitter phase perturbations}, \ref{phantom phase perturbations} and \ref{frw phase perturbations}), 
we find expressions for $v_{k}$ in the three stages of the model. The unknown coefficients
are determined in  \ref{Unknown coefficients}, and the scalar and tensor power spectra are computed in \ref{scalar power spectrum} and \ref{tensor power spectrum} respectively.
\subsection{$v_{k}$ in the de Sitter Phase }
\label{de sitter phase perturbations}
For a scalar field in de Sitter space, the gauge-invariant Mukhanov variable $v_{k}(\eta)$ can be
shown to satisfy the equation (see, for instance, \cite{langlois})
\begin{equation}
\label{ads}
v_{k}''(\eta)+\left[ k^{2}-\frac{a''(\eta)}{a(\eta)}\right] v_{k}(\eta) = 0
\end{equation}
The solutions to this equation are the Bunch-Davies mode functions:
\begin{equation}
v_{k}(\eta) = \frac{e^{-ik\eta}}{\sqrt{2k}}\left( 1-\frac{i}{k\eta}\right)
\end{equation}


\subsection{$v_{k}$ in the Phantom Phase }
\label{phantom phase perturbations}
In this case we will calculate $v_{k}$ starting from first principles.
\subsubsection{Matter and metric perturbations}
\label{phmatterandmetricpertubations}
The first step is to perturb the matter and metric. Working in longitudinal gauge and assuming no 
anisotropic stress, the scalar metric perturbations are written as:
\begin{equation}
g_{\mu\nu}=a^{2}\left(
\begin {array}{cccc}
1+2\Phi &0&0&0\\
0&-1+2\Phi &0&0\\
0&0&-1+2\Phi &0\\
0&0&0&-1+2\Phi
\end {array}
\right)
\end{equation}
Given our choice of gauge, the metric perturbation $\Phi(\eta, \vec{x})$ coincides with the gauge-invariant Bardeen potential (see, for example \cite{Mukhanov}). 

The phantom matter field $\phi_{p}(\eta, \vec{x})$ is perturbed as follows:
\begin{equation}
\label{pertKG}
\phi_{p}(\eta, \vec{x})=\phi(\eta)+\delta\phi(\vec{x},\eta)
\end{equation}

\subsubsection{Evolution of the perturbations}
\label{evolutionoftheperturbations}
To find time evolution of the perturbations, we use the perturbed Einstein equations up to first order.
The i-i component of the zero-th order equations reads:
\begin{equation}
\label{zeroorderspace}
-\Lambda a^{2}+\left( \frac{a'}{a}\right)^{2}-2\frac{a''}{a}=8\pi G\left[\lambda \frac{\phi'^{2}}{2}
-a^{2} V(\phi)\right]\\
\end{equation}
while the 0-0 component reads:
\begin{equation}
\label{zeroordertime}
\Lambda a^{2}+3\left( \frac{a'}{a}\right)^{2}=8\pi G\left[\lambda \frac{\phi'^{2}}{2}
+a^{2} V(\phi)\right]
\end{equation}
Adding these equations, one obtains the familiar relationship:
\begin{eqnarray}
\label{familiar}{\cal H}^{2}-{\cal H}'=4\pi G \lambda\phi'^{2}
\end{eqnarray}
The i-i, 0-0 and 0-i components of the first order equations are respectively:
\begin{align}
&\Lambda a^{2}\Phi-2\Phi\left(\frac{a'}{a}\right)^{2}+4\Phi\frac{a''}{a}+3\frac{a'}{a}\Phi' +\Phi'' =&\nonumber\\
\label{firstorderspace} &8\pi G\left[a^{2}V(\phi)\Phi - \frac{1}{2}a^{2}V_{\phi}\delta\phi+\frac{1}{2}\lambda\phi'\delta\phi'-\lambda\Phi\phi'^{2} \right]\\
&\Lambda a^{2}\Phi-3\frac{a'}{a}\Phi' +\nabla^{2}\Phi =&\nonumber\\
\label{firstordertime}   &8\pi G\left[a^{2}V(\phi)\Phi + \frac{1}{2}a^{2}V_{\phi}\delta\phi+\frac{1}{2}\lambda\phi'\delta\phi' \right]&\\
\label{constraint}&\Phi'+\frac{a'}{a}\Phi = 4\pi G\lambda\phi'\delta\phi&
\end{align}
(where $V_{\phi}=\partial V/\partial \phi$).

Perturbing the Klein-Gordon equation Eq.~\eqref{KG} using Eq.~\eqref{pertKG} yields, at zero-th order:
\begin{equation}
\label{phantom}
\lambda\left[\phi''+2{\cal H}\phi'\right]=-a^{2}V_{\phi}
\end{equation}
Using Eq.~\eqref{familiar}, Eq.~\eqref{firstorderspace}, Eq.~\eqref{firstordertime}, Eq.~\eqref{constraint} and Eq.~\eqref{phantom}, one obtains 
the equation of motion of $\Phi(\eta, \vec{x})$:
\begin{equation}
\Phi''-\nabla^{2}\Phi+2\left({\cal H}-\frac{\phi''}{\phi'}\right)\Phi'+2\left({\cal H}'
-{\cal H}\frac{\phi''}{\phi'}\right)\Phi=0
\end{equation}
Applying the Fourier transform:
\begin{equation}
\Phi( \vec{x},\eta)=\int \frac{d^{3}\vec{k}}{(2\pi)^{3/2}}\,\Phi_{k}(\eta)\, e^{i \vec{k}.\vec{x}}
\end{equation}
we obtain:
\begin{equation}
\label{master}
\Phi''_{k}+2\left({\cal H}-\frac{\phi''}{\phi'}\right)\Phi'_{k}+\left[k^{2}+2\left({\cal H}'
-{\cal H}\frac{\phi''}{\phi'}\right)\right]\Phi_{k}=0
\end{equation}

Note that Eq.~\eqref{master} is independent of $\lambda$, indicating that it has the same form for a phantom field as 
it would for a normal field. This is a surprising result, since all the equations used to derive Eq.~\eqref{master} are $\lambda$ dependent. Physically, this result implies that the evolution of the metric perturbation is insensitive to whether the matter content of the Universe is normal or phantom. 

To solve Eq.~\eqref{master}, we need to determine the dynamics of the phantom field $\phi(\eta)$, which is in turn determined by the potential $V(\phi)$. We choose a particular form of the potential which allows for a solution in closed form:

\begin{equation}
\label{potential}
V(\phi_{p})=K^{2}\frac{(3+\alpha)}{2\alpha}\,\exp\left[\frac{2\alpha H_{\Lambda}}{K}\phi_{p}\right]
\end{equation} 
 Here $K$ is a constant that has the dimensions of mass squared, and sets the scale of the potential.

For this potential, it is easy to verify that the exact form of $\phi(\eta)$ which satisfies Eq.~\eqref{phantom} (with of course $\lambda=-1$ as is the case in the phantom phase) is: 
\begin{equation}
\label{phisolution}
\phi(\eta)=-\frac{K}{H_{\Lambda}(1+\alpha)}\,\ln\left[-\alpha-H_{\Lambda}(1+\alpha)\eta\right]
\end{equation}
Further, to facilitate the back-reaction as discussed in Section \ref{The Phantom Phase}, and satisfy our ansatz given by Eq.~\eqref{alphadef}
we must require that our field satisfies the Friedmann equation Eq.~\eqref{zeroordertime}. The result of this is to fix the value of $K$:
 \begin{equation}
 \label{K}
 K=H_{\Lambda}\sqrt{\frac{3\alpha }{4\pi G}}
 \end{equation}
 The matter field $\phi(\eta)$ has the interesting property that 
 \begin{equation}
 \label{interestingproperty}
 \frac{\phi''(\eta)}{\phi'(\eta)}=\left(1+\alpha\right){\cal H}(\eta)
 \end{equation}
Using Eq.~\eqref{interestingproperty}, Eq.~\eqref{master} reduces to 
\begin{equation}
\Phi_{k}''-2{\cal H}\alpha\Phi_{k}'+\left[k^{2}-2{\cal H}^{2}\alpha\right]\Phi_{k}=0
\end{equation}
This is a familiar second order differential equation of the form:
\begin{equation}
\Phi_{k}''+P(\eta)\Phi_{k}'+Q(\eta)\Phi_{k}=0
\end{equation}
with
\begin{eqnarray*}
P(\eta)&=&-2{\cal H}(\eta)\alpha\\
Q(\eta)&= &k^{2}-2{\cal H}^{2}(\eta)\alpha
\end{eqnarray*}
which has the solution (see e.g. \cite{wiesstein})
\begin{equation}
\label{gensol}
\Phi_{k}(\eta)=e^{-\frac{1}{2}\int P(\eta) \,d\eta}\chi(\eta)=a^{\alpha}\chi(\eta)
\end{equation}
where $\chi(\eta)$ satisfies the differential equation
\begin{equation}
\chi''(\eta)+\left[Q-\frac{1}{2}P'-\frac{1}{4}P^{2}\right]\chi(\eta)=0
\end{equation}
In our case, this reduces to 
\begin{equation}
\label{chiineta}
\chi''(\eta)+\left[k^{2}+\frac{\alpha H_{\Lambda}^{2}}{\left[\alpha+H_{\Lambda}(1+\alpha)\eta\right]^{2}}\right]\chi(\eta)=0
\end{equation}
At this point it is convenient to temporarily switch to a new time variable $x$ defined by
\begin{equation}\label{x}
x=\alpha+H_{\Lambda}(1+\alpha)\eta
\end{equation}
Note that $x=-1$ when $\eta=\eta_{i}$ and from Eq.~\eqref{eta_f}, $x=O\left[H_{\Lambda}/H_{f}\right]$ or $x\simeq0$ when $\eta=\eta_{f}$.

With $x$ as the time variable, Eq.~\eqref{chiineta} takes the form
\begin{equation}
\label{chiinx}
\frac{d^{2}\chi(x)}{dx^{2}}+\left[m^{2}-\left(p^{2}-\frac{1}{4}\right)\frac{1}{x^{2}}\right]\chi(x)=0
\end{equation}
where $m$ and $p$ are defined by
\begin{align}
\label{m}m&=\frac{k}{H_{\Lambda}(1+\alpha)}&\\
\label{p}p&=\left(\frac{1}{4}-\frac{\alpha}{(\alpha+1)^{2}}\right)^{1/2}&
\end{align}
The solution to Eq.~\eqref{chiinx} can be written in terms of Bessel functions as 
\begin{equation}
\label{chisol}
\chi(x)=\sqrt{mx}\left[A_{m}J_{p}(mx)+B_{m}Y_{p}(mx)\right]
\end{equation}
where $A_{m}$, $B_{m}$ are constants and $J_{p}$ and $Y_{p}$ denote Bessel functions of the first and second kind of order $p$ respectively.

Putting together Eq.~\eqref{chisol} and Eq.~\eqref{gensol}, we conclude that the time evolution of $\Phi_{k}$ is fully described by the equation
\begin{equation}
\label{phifinal}
\Phi_{k}(\eta)=a^{\alpha}\sqrt{mx}\left[A_{m}J_{p}(mx)+B_{m}Y_{p}(mx)\right]
\end{equation}
where $m$, $x$ and $p$ are defined by Eq.~\eqref{m}, Eq.~\eqref{x} and Eq.~\eqref{p} respectively, and $\alpha$ is defined in Eq.~\eqref{alphadef}.

\subsubsection{Calculating the Mukhanov Variable}
\label{calculatingthemukhanovvariable}
The Mukhanov variable $v_{k}$ can also be defined by the relationship (see, for example \cite{brandenberger1}):
\begin{equation}
\label{v}
v_{k}={\cal R}_{k}z
\end{equation}
Where ${\cal R}_{k}$ is the curvature perturbation on co-moving hyper-surfaces at length scale $k$. 

We are first going to compute ${\cal R}_{k}$ and then $z$ and then use Eq.~\eqref{v} to compute $v_{k}$.
\paragraph{Calculation of ${\cal R}_{k}$: }
The curvature perturbation on co-moving hyper-surfaces (${\cal R}_{k}$) is defined by the relation (in momentum space) \cite{riotto}:
\begin{equation}
\label{R}
{\cal R}_{k}=\Phi_{k}+{\cal H}\frac{\delta\phi}{\phi'}
\end{equation}
Noticing that for this space, we have 
\begin{equation}
\label{alpha}
\frac{{\cal H'}}{{\cal H}^{2}}-1=\alpha
\end{equation}
and using  equations Eq.~\eqref{constraint} and Eq.~\eqref{familiar}, we can eliminate the $\delta\phi$ in Eq.~\eqref{R}, giving us an
expression for ${\cal R}_{k}$ involving $\Phi_{k}$ and $\Phi'_{k}$ as the only first order variables:
\begin{equation}
\label{newR}
{\cal R}_{k}=\Phi_{k}-\frac{1}{\alpha {\cal H}}\left[ \Phi'_{k}+{\cal H}\Phi_{k}\right]
\end{equation}
(Again, the absence of $\lambda$ indicates that the expression for ${\cal R}_{k}$ is insensitive to
 whether the field is real or phantom.)
 \paragraph{Calculation of $z$:}
 In the definition of $z$ Eq.~\eqref{z} we eliminate $\phi'$ using Eq.~\eqref{familiar}, and introduce $\alpha$ 
using Eq.~\eqref{alpha} to get
\begin{eqnarray}
z&=&a\sqrt{\frac{\alpha}{-4\pi G \lambda}}\nonumber\\
\label{newz}&=&a m_{\text{Pl}}\sqrt{\frac{2\alpha}{-\lambda}}
\end{eqnarray}
where $m_{\text{Pl}}$ is the reduced Planck mass defined by $m_{\text{Pl}}=\sqrt{1/8\pi G}$, where $G$ is Newton's gravitational constant.
\paragraph{Final expression for $v_{k}$:}
 Combining Eq.~\eqref{v}, Eq.~\eqref{newR} and Eq.~\eqref{newz}, we can express $v_{k}$ entirely in terms of $\Phi_{k}$ and 
 $\Phi'_{k}$ as follows:
 \begin{equation}
v_{k}= a\sqrt{\frac{\alpha}{-4\pi G \lambda}}\left[ \Phi_{k}-\frac{1}{\alpha {\cal H}}\left( \Phi'_{k}+{\cal H}\Phi_{k}\right)\right]
 \end{equation}

\subsection{$v_{k}$ in the FRW Phase}
 \label{frw phase perturbations}
 Here we have a gauge field in a geometry described by Eq.~\eqref{afrw}. From the theory of cosmological perturbations, 
we know that $v_{k}$ satisfies \cite{langlois} 
\begin{equation}
\label{afrwz}
v_{k}''(\eta)+\left[ k^{2}-\frac{z''(\eta)}{z(\eta)}\right] v_{k}(\eta) = 0
\end{equation}
But since $z(\eta)=2m_{\text{Pl}}a(\eta)$ in this space, the above equation simply reduces to 
Eq.~\eqref{ads}. Solving Eq.~\eqref{ads} for this space, we find that $v_{k}$ is described by the solution:
 \begin{equation}
 \label{vfrw}
 v_{k}(\tau)=\alpha_{k}e^{ik\tau}+\beta_{k}e^{-ik\tau}
 \end{equation}
 where $\alpha_{k}$ and $\beta_{k}$ are constants of integration and $\tau=\eta-\eta_{f}>0$

\subsection{Calculation of Unknown Constants}
\label{Unknown coefficients}

The above calculations produced four unknown constants $A_{m}$, $B_{m}$, in Eq.~\eqref{chisol} and $\alpha_{k}$ 
and $\beta_{k}$, in Eq.~\eqref{vfrw}. These constants can be determined by demanding continuity of $v_{k}$ and its time derivative at the two transition times $\eta_{i}$ and $\eta_{f}$. 

To determine $A_{m}$ and $B_{m}$, we perform the above matching process at $\eta=\eta_{i}=-\frac{1}{H_{\Lambda}}$, 
or in terms of $x$, (from Eq.~\eqref{x}), at $x=x_{i}=-1$ . 
In other words, we need to simultaneously solve the equations
\begin{eqnarray}
v_{k}\text{(de Sitter)}\vert_{\left(\eta=-1/H_{\Lambda}\right)}=v_{k}\text{(phantom)}\vert_{(x=-1)}\nonumber\\
\label{etai}v_{k}'\text{(de Sitter)}\vert_{\left(\eta=-1/H_{\Lambda}\right)}=v_{k}'\text{(phantom)}\vert_{(x=-1)}
\end{eqnarray}
to find the unknowns $A_{m}$ and $B_{m}$.

The expression obtained for 
$v_{k}$ in the phantom phase (by substituting the values of the above constants) is fairly complicated. However, since we are only interested in the super-horizon modes, we can make the approximation that $k\text{ (or $m$)}\rightarrow0$, and use the appropriate asymptotic forms of the Bessel functions $J_{p}(mx)$ and $Y_{p}(mx)$. Also, since $\alpha$ is large, from Eq.~\eqref{p}, 
$p\simeq\frac{1}{2}$. With these simplifications the expression for $v_{k}$ (in the phantom phase) reduces to:
\begin{eqnarray}
\label{vk}v_{k}(x)&\simeq&-\frac{i\,e^{ik/H_{\Lambda}}a^{(1+\alpha)}H_{\Lambda}(-1+x\alpha)}{\sqrt{2}k^{3/2}(1+\alpha )}\nonumber\\
& &+O\left[k^{-1/2}\right]\\
\label{vk'}v_{k}'(x)&\simeq&-\frac{i\,e^{ik/H_{\Lambda}}a^{(1+\alpha)}H_{\Lambda}^{2}}{\sqrt{2}k^{3/2}x}
+O\left[k^{-1/2}\right]
\end{eqnarray}

Now having fully determined the form of $v_{k}(\eta)$ during the phantom phase, 
we can determine the coefficients $\alpha_{k}$ and $\beta_{k}$ in Eq.~\eqref{vfrw} to determine $v_{k}(\eta)$ in the final FRW phase. In particular, we need to solve
simultaneously the equations:
\begin{eqnarray}
v_{k}\text{(phantom)}\vert_{\left(x\simeq0\right)}=v_{k}\text{(FRW)}\vert_{(\eta=\eta_{f})}\nonumber\\
\label{etaf}v_{k}'\text{(phantom)}\vert_{\left(x\simeq0\right)}=v_{k}'\text{(FRW)}\vert_{(\eta=\eta_{f})}
\end{eqnarray}
For brevity, let us call the leading term in $k$ in the expression for $v_{k}$  (Eq.~\eqref{vk}) at $\eta=\eta_{f}$ (or $x\simeq0$)
 as $l_{1}$ and the leading term in $k$ in the expression for 
$v_{k}'$  (Eq.~\eqref{vk'}) at $x\simeq0$ as $l_{2}$. Thus we have 
\begin{equation}
\label{l1}l_{1}=\frac{i\,e^{ik/H_{\Lambda}}a^{(1+\alpha)}H_{\Lambda}}{\sqrt{2}k^{3/2}(1+\alpha )}
\end{equation}
and,
\begin{eqnarray}
\label{l2}l_{2}&=&-\frac{i\,e^{ik/H_{\Lambda}}a^{(1+\alpha)}H_{\Lambda}^{2}}{\sqrt{2}k^{3/2}x_{f}}\nonumber\\
&=&\frac{i\,e^{i/kH_{\Lambda}}a_{f}^{(2+2\alpha)}H_{\Lambda}^{2}}{\sqrt{2}k^{3/2}}
\end{eqnarray}
Where in the last manipulation we have used the result that $-x_{f}=a_{f}^{-(1+\alpha)}$, which follows from the form 
of the scale factor during the phantom phase (Eq.~\eqref{scalefactor}) and the definition of $x$ (Eq.~\eqref{x}). Eq.~\eqref{etaf} now implies that
\begin{eqnarray}
\alpha_{k}&=\frac{1}{2}\left[l_{1}+\frac{l_{2}}{ik}\right]&\nonumber\\
\label{alphabeta}\beta_{k}&=\frac{1}{2}\left[l_{1}-\frac{l_{2}}{ik}\right]
\end{eqnarray}
Using Eq.~\eqref{H} at ($\eta=\eta_{f}$) to find
\begin{equation}
\label{HiHf}
a_{f}^{\alpha}H_{\Lambda}=H_{f} 
\end{equation}
we note from equations Eq.~\eqref{l2} that,  
\begin{equation}
\frac{l_{1}}{l_{2}}=\frac{1}{(1+\alpha)a_{f}^{1+\alpha}H_{\Lambda}}=\frac{1}{a_{f}H_{f}(1+\alpha)}\ll1
\end{equation}
since both $\alpha$ and $H_{f}$ are large. Eq.~\eqref{alphabeta} now reduces to 
\begin{eqnarray}
\alpha_{k}&\simeq&\quad\frac{1}{2}\left[\frac{l_{2}}{ik}\right]\nonumber\\
\beta_{k}&\simeq&\,-\frac{1}{2}\left[\frac{l_{2}}{ik}\right]
\end{eqnarray}
Hence the form of $v_{k}(\eta)$ in the phantom phase becomes:
\begin{eqnarray}
\label{vfinalfrw}v_{k}(\eta) &\simeq&\frac{l_{2}}{2ik}e^{ik\tau}-\frac{l_{2}}{2ik}e^{-ik\tau}\nonumber\\
                             &=&l_{2}\left(\frac{sin(k\tau)}{k\tau}\right)\tau
\end{eqnarray}

\subsection{Determination of the Scalar Power Spectrum}
\label{scalar power spectrum}
Now we are in a position to determine the power spectrum of the co-moving curvature perturbation in the 
FRW space using Eq.~\eqref{powerspectrum} which gives
\begin{equation}
{\cal P}_{\cal R}=\frac{k^{3}}{2\pi^{2}}\left| l_{2}\frac{sin(k\tau)}{k\tau}\frac{\tau}{z_{\text{FRW}}(\eta)}\right|^{2}
\end{equation}

Making the approximation $a_{\text{FRW}}(\eta)=a_{f}^{2}H_{f}\tau$ from Eq.~\eqref{afrw} since at the time a mode re-enters the horizon, $\eta\gg\eta_{f}$, using the 
relation Eq.~\eqref{HiHf}, and taking the limit $k\rightarrow0$, we finally obtain
\begin{eqnarray}
{\cal P}_{\cal R}&\simeq&\frac{k^{3}}{2\pi^{2}}\left|\frac{i\,e^{-ik/H_{\Lambda}}}{2\sqrt{2}k^{3/2}m_{\text{Pl}}}H_{f}\right|^{2}\\
                 &  =   &\left(\frac{H_{f}}{4\pi m_{\text{Pl}}}\right)^{2}
\end{eqnarray}
Hence we find that our model produces a (nearly) scale invariant spectrum of cosmological perturbations, with amplitude set by 
$H_{f}/m_{\text{Pl}}$. If we assume $H_{f}\sim 10^{14}\text{ GeV}$ (approximately GUT scale), then the power spectrum matches the COBE DMR observations \cite{Smoot} of CMB temperature fluctuations of order $10^{-5}$. In other words, the perturbation spectrum can have an amplitude sufficiently large to seed the cosmological structure that we see today.

\subsection{Determination of the Tensor Power Spectrum}
\label{tensor power spectrum}

We know from the theory of cosmological perturbations (see, for example, \cite{langlois}) that gravitational waves are essentially equivalent to two massless scalar fields (for each polarization) up to a renormalization factor of $2/m_{\text{Pl}}$. Hence we can write the tensor power spectrum ${\cal P}_{{\cal R}T}$ as
\begin{equation}
{\cal P}_{{\cal R}T}=2\frac{4}{m_{\text{Pl}}^{2}}{\cal P}_{{\cal R}\psi}
\end{equation}
where the first factor on the right comes from the two polarization states, the second represents the renormalization mentioned above, and ${\cal P}_{{\cal R}\psi}$ is the spectrum of perturbations of a massless scalar field $\psi$ other than, and not interacting with, the NEC-violating field. The $\psi$-field perturbations can be computed using essentially the same machinery as above, with a small difference in the final step:
\begin{enumerate}
\item We solve Eq.~\eqref{master} with $\psi$ representing the matter field. 
\item We find expressions for $v_{k}$ in the three stages of the model (up to constants of integration).
\item We demand the continuity of $v_{k}$ and $v_{k}'$ at $\eta_{i}$ and $\eta_{f}$ to fully specify the expression for $v_{k}$ in the FRW region (that is, evaluate the undetermined constants obtained in the previous step), and from this, compute the perturbation spectrum ${\cal P}_{{\cal R}\psi}$ using the relation
\begin{equation}
\label{tensor powerspectrum}{\cal P}_{{\cal R}\psi}(\eta )=\frac{k^{3}}{2\pi^{2}}\left\vert\frac{v_{k}(\eta )}{a(\eta)}\right\vert_{\text{FRW}} ^{2}
\end{equation}
\end{enumerate}

The final expression for the power spectrum of tensor perturbations turns out to be
\begin{equation}
\label{Tensor Final Expression}
{\cal P}_{{\cal R}T}=8\left(\frac{H_{\Lambda}}{a_{f}^{3}\pi m_{\text{Pl}}}\right)^{2}\simeq8\left(\frac{H_{\Lambda}}{\pi m_{\text{Pl}}}\right)^{2}
\end{equation}
The last step follows because it is easy to show (by Taylor expanding $a(\eta)$ about the point $\eta=\eta_{i}$) that the scale factor hardly changes from its initial value of $1$ during the NEC-violating event.

This result agrees (up to $O(1)$ numerical factors) with the corresponding result obtained in \cite{islands}. 

A similar scenario was investigated by Y.S Piao \cite{piao}, and while he obtains a scale invariant spectrum of scalar perturbations, the tensor perturbation spectrum in his calculation turns out to be blue-shifted. The source of the discrepancy could lie in his assumption that the canonical relationship satisfied by $v_{k}$ in the case of the tensor spectrum has a time-dependent mass given by $a''(\eta)/a(\eta)$ (Eq.~(12) in \cite{piao}) during the phantom phase. In our approach we derive $v_{k}$ from first principles and find that the time dependent mass can have a more complicated form.

\section{Conclusions}
\label{Conclusions}
To summarize, in this paper we have investigated a cosmological model in which islands of matter are created from 
NEC-violating explosive events in a cosmological constant dominated inflating Universe. Our approach was to model the behavior of the matter field during the NEC-violating fluctuation as a classical phantom field. While an ideal approach would be a full quantum mechanical calculation, it is hoped that our classical calculation can capture the essential elements of the physics involved.

Our calculations yield, (using an exactly solvable potential), an adiabatic spectrum of scale-invariant perturbations, whose amplitude is determined by the value of the Hubble constant at the end of the NEC-violating fluctuation. If we assume the latter to be approximately GUT scale, the perturbation spectrum turns out to have an amplitude sufficient to seed the cosmological structure seen today. We also obtain a scale-invariant spectrum of gravitational waves of amplitude set by $(H_{\Lambda}/m_{\text{Pl}})$.

The essential elements of Island Cosmology, namely, the $\Lambda$-sea and the scale invariant perturbation spectra, are supported by current observations. The NEC-violating quantum fluctuations which create the matter islands are a clear consequence of the basic formalism of quantum field theory. Island cosmology therefore holds out the tantalizing notion that the observed Universe could have had a purely quantum field theoretic origin. This paper lends further support to this idea.

\begin{acknowledgments} 
I am deeply indebted to Tanmay Vachaspati for guidance and discussions. I would also like to thank Francesc Ferrer, Irit Maor and Dejan Stojkovich for feedback, and the Michigan Center for Theoretical Physics for hospitality. This work was supported by the U.S. Department of Energy and NASA.
\end{acknowledgments}

\end{document}